# Room-temperature anisotropic plasma mirror and polarization-controlled optical switch based on Type-II Weyl semimetal $WP_2$

Kaixuan Zhang,[1,†] Yongping Du,[2,†] Zeming Qi,[3,†] Bin Cheng,[1] Xiaodong Fan,[1] Laiming Wei,[1,*] Lin Li,[1] Dongli Wang,[1] Guolin Yu,[4] Shuhong Hu,[4] Changhong Sun,[4] Zhiming Huang,[4] Junhao Chu,[4] Xiangang Wan,[5] and Changgan Zeng[1,*]

[1]*International Center for Quantum Design of Functional Materials, Hefei National Laboratory for Physical Sciences at the Microscale, CAS Key Laboratory of Strongly Coupled Quantum Matter Physics, Department of Physics, and Synergetic Innovation Center of Quantum Information and Quantum Physics, University of Science and Technology of China, Hefei, Anhui 230026, China*

[2]*Department of Applied Physics and Institution of Energy and Microstructure, Nanjing University of Science and Technology, Nanjing, Jiangsu 210094, China*

[3]*National Synchrotron Radiation Laboratory, University of Science and Technology of China, Hefei, Anhui 230029, China*

[4]*National Laboratory for Infrared Physics, Shanghai Institute of Technical Physics, Chinese Academy of Sciences, Shanghai 200083, China*

[5]*National Laboratory of Solid State Microstructures, Department of Physics, and Collaborative Innovation Center of Advanced Microstructures, Nanjing University, Nanjing, Jiangsu 210093, China*

Anisotropy in electronic structures may ignite intriguing anisotropic optical responses, as well demonstrated in various systems including superconductors, semiconductors and even topological Weyl semimetals. Meanwhile, it is well established in metal optics that the metal reflectance declines from one to zero when the photon frequency is above the plasma frequency $\omega_p$, behaving as a plasma mirror. However, the exploration of anisotropic plasma mirrors and corresponding applications remains elusive, especially at room temperature. Here, we discover a pronounced anisotropic plasma reflectance edge in the type-II Weyl semimetal $WP_2$, with an anisotropy ratio of $\omega_p$ up to 1.5. Such anisotropic plasma mirror behavior and its robustness against temperature promise optical device applications over a wide temperature range. For example, the high sensitivity of polarization-resolved plasma reflectance edge renders $WP_2$ an inherent polarization detector. We further achieve a room-temperature $WP_2$-based optical switch, effectively controlled by simply tuning the light polarization. These findings extend the frontiers of metal optics as a discipline and promise the design of multifunctional devices combining both topological and optical features.

## I. INTRODUCTION

Anisotropies in atomic and electronic structures have been discovered to spark an array of intriguing physical phenomena, including anisotropic optical responses [1–12]. For example, anisotropic optical conductivity was revealed in the parent compounds of iron arsenide superconductors, arising from the anisotropic energy gap opening [1]. Anisotropic optical absorption and photoluminescence were also discovered in two-dimensional black phosphorus semiconductor, and were attributed to the anisotropies in selection rule and effective mass [3–5]. Recently, anisotropic photocurrent responses were unveiled in Weyl semimetals due to the chirality selection rule and asymmetric Pauli blockade in finitely tilted Weyl cones [8, 9].

On the other hand, metals can be regarded as plasma mirrors, with the reflectance edge determined by the plasma frequency $\omega_p = \sqrt{ne^2/\epsilon_0 m^*}$ [13, 14], where $n$ is the carrier density, $e$ is the elementary charge, $\epsilon_0$ is the vacuum permittivity, and $m^*$ is the effective mass. Usually, metals possess nearly isotropic Fermi surfaces and the corresponding isotropic plasma reflectance edge [13, 14]. In contrast, some semimetals possess highly anisotropic Fermi surfaces, such as bismuth [15, 16] and $WTe_2$ [17–20], which are expected to exhibit anisotropic plasma edges. These semimetals however possess low $\omega_p$, thus preventing the achievement of anisotropic plasma mirrors at high temperatures. Recently, $WP_2$ was theoretically predicted to be a robust new Type-II Weyl semimetal [21] with highly anisotropic Fermi surfaces [21–23]. Herein, we describe the discovery of an anisotropic plasma reflectance edge in $WP_2$, which is pronounced even at room temperature, and further demonstrate a typical applica-tion of a polarization-controlled optical switch.

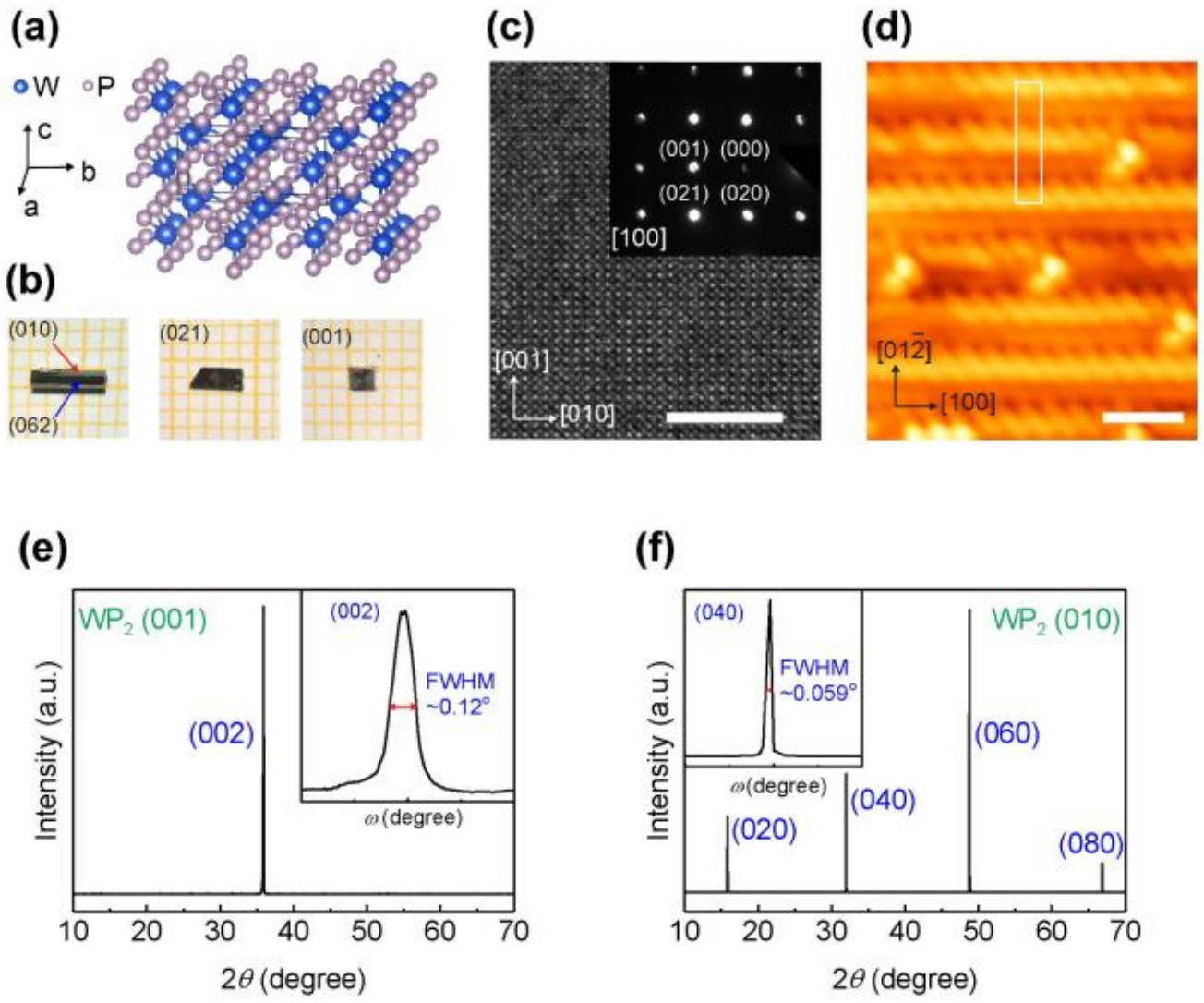

FIG. 1. Structure characterizations of $WP_2$. (a) Crystallographic structure. The black lines indicate a unit cell. (b) Optical images of the samples with different crystalline surfaces adopted for XRD and optical reflectance measurements. The orange squares in the background are $1 \times 1$ mm$^2$. (c) TEM and selected area electron diffraction images along the [100] direction. The scale bar is 6 nm. (d) Atomic-resolution STM image of the cleaved (021) surface. The white rectangle denotes a surface unit cell. The scale bar is 1 nm. (e,f) XRD patterns on the (001) (e) and (010) (f) surfaces, respectively. The insets show the rocking curves with a small full width at half maximum (FWHM).

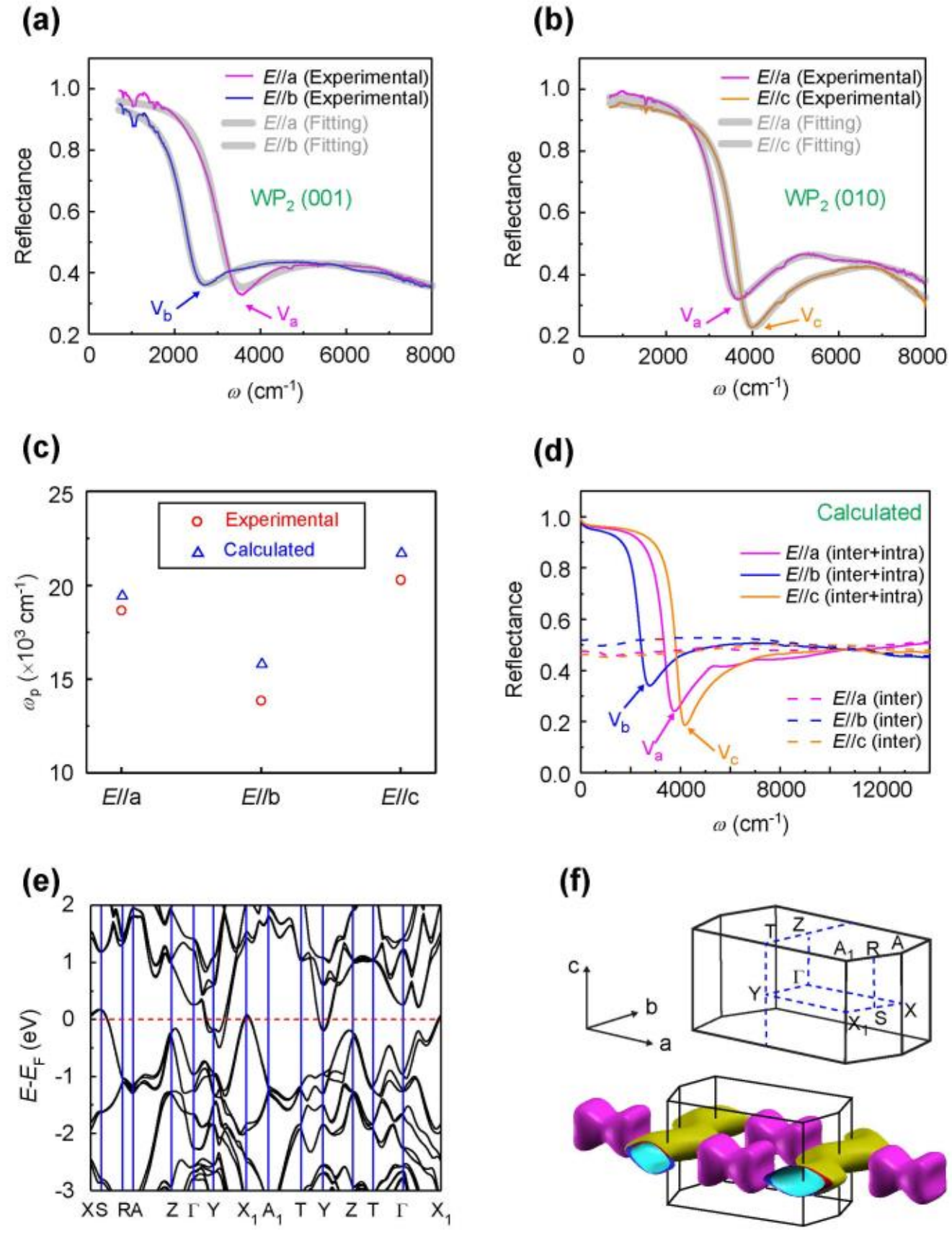

FIG. 2. Anisotropic reflectance and electronic structures of $WP_2$. (a,b) Measured reflectance spectra of the (001) (a) and (010) (b) surfaces, respectively. The light polarization $E$ in (a) is parallel to the a- (pink) and b-axes (blue), whereas $E$ in (b) is parallel to the a- (pink) and c-axes (yellow). The pink, blue, and yellow arrows indicate the reflectance valleys for $E//$a ($V_a$), $E//$b ($V_b$), and $E//$c ($V_c$), respectively. The gray curves are the fitting curves based on the two-Drude model. (c) $\omega_p$ extracted from fitting the experimental data (circles) and from theoretical calculations (triangles), respectively. (d) Calculated reflectance spectra for $E//$a (pink), $E//$b (blue), and $E//$c (yellow), respectively. The solid curves take into account both the intraband and interband excitations, whereas the dashed curves consider only interband excitations. (e) Calculated band structures along high symmetry directions. (f) Calculated Fermi surfaces. The bow tie-shaped closed pockets are electron Fermi surfaces, while the spaghetti-shaped open pockets are hole Fermi surfaces.

## II. RESULTS AND DISCUSSION

### A. Structure characterization of $WP_2$

$WP_2$ single crystals with orthorhombic structure [$\beta$-phase, Fig. 1(a)] were grown via chemical vapor transport [24] (see more details in the Supplemental Material (SM) [25]). Figure 1(b) depicts several natural crystal faces including the (010), (062), and (021) surfaces. The high quality of the $WP_2$ single crystals at atomic and macroscopic scales are demonstrated by transmission electron microscopy (TEM) image [Fig. 1(c)], scanning tunneling microscopy (STM) image [Fig. 1(d)], and X-ray diffraction (XRD) patterns [Figs. 1(e) and 1(f)].

### B. Room-temperature anisotropic plasma mirror behavior of $WP_2$

Polarization-resolved reflectance spectra were measured by a Fourier-transform infrared spectrometry. The light polarization $E$ was rotated in the (001) plane to measure the reflectance spectra for $E$ parallel to the crystallographic a- and b-axes, and was rotated in the (010) plane to measure the spectra for $E$ along the a- and c-axes. Figures 2(a) and 2(b) show a well-defined sharp reflectance edge emerging in these spectra, followed by a reflectance valley denoted as $V_a$ at $\sim$ 3600 cm$^{-1}$ for $E//$a, $V_b$ at $\sim$ 2700 cm$^{-1}$ for $E//$b, and $V_c$ at $\sim$ 4000 cm$^{-1}$ for $E//$c, respectively. A reflectance valley typically develops near the screened plasma frequency $\omega_p^* = \omega_p/\sqrt{\epsilon_\infty}$, where $\epsilon_\infty$ is the permittivity at high frequency [16]. Therefore, the varied valley wavenumbers for $E//$a, $E//$b, and $E//$c reflect anisotropy in the plasma frequency.

Quantitative analysis of the plasma frequency is accomplished by using the RefFIT program [26] to fit the reflectance curves according to the two-Drude model [19] with the complex dielectric function:

$$\epsilon(\omega) \;=\; \epsilon_\infty - \sum_{j=1}^{2} \frac{\omega_{p,j}^2}{\omega^2 + i\omega/\tau_j} + \sum_k \frac{\Omega_{p,k}^2}{\omega_{0,k}^2 - \omega^2 - i\omega\gamma_k}$$

where $\omega_{p,j}$ are the free carrier plasma frequencies for electrons and holes, $\tau_j$ are the free carrier scattering times for electrons and holes, $\Omega_{p,k}$ are the oscillator strengths for phonons and interband electronic transitions, $\omega_{0,k}$ are the phonon and interband transition frequencies, and $\gamma_k$ is the width of the corresponding transition (see more details in Table S1). The bare plasma frequencies [20] for $E//$a, $E//$b, and $E//$c are $\omega_{p,a}^2 = \omega_{p,a,1}^2 + \omega_{p,a,2}^2$, $\omega_{p,b}^2 = \omega_{p,b,1}^2 + \omega_{p,b,2}^2$, and $\omega_{p,c}^2 = \omega_{p,c,1}^2 + \omega_{p,c,2}^2$, respectively. The fitting curves for $E//$a, $E//$b, and $E//$c are plotted with thick gray curves in Figs. 2(a) and 2(b). The $\omega_p$ values estimated from the two-Drude fitting model are displayed in Fig. 2c (circles), which illustrate strong anisotropies for $E//$a, $E//$b, and $E//$c. For example, $\omega_{p,a}/\omega_{p,b}$ and $\omega_{p,c}/\omega_{p,b}$ are about 1.35 and 1.46, consistent with the valley wavenumber ratios $V_a$:$V_b$ ($\sim$ 1.33) and $V_c$:$V_b$ ($\sim$ 1.48), respectively.

The sharp reflectance edge revealed here suggests that the interband excitations are well separated from the plasma edge. By contrast, some metals such as copper exhibit smeared reflectance edges that is considerably mixed with the pronounced interband electronic transition [13]. To reinforce this point, theoretical calculations were performed (see more details in the SM [25]). When only interband excitations are considered, a clear reflectance edge is absent in the calculated reflectance spectra (see dashed curves in Fig. 2(d)). When the intraband excitations are taken into account, the calculated spectra exhibit sharp and anisotropic reflectance edges, with the reflectance valley wavenumbers approximating those of experimental measurements (see solid curves in Fig. 2(d)).

Figure 2(c) shows that the $\omega_p$ values extracted by fitting the experimental data (circles), are consistent with theoretical predictions (triangles). The effective mass, which is extracted from the formula $\omega_p^2 = ne^2/\epsilon_0 m^*$, reaches its maximum along the b-axis with a mass anisotropy ratio $\eta_{ab} = m_b^*/m_a^* = \omega_{p,a}^2/\omega_{p,b}^2 \sim 1.8$, which is qualitatively comparable to the calculated value of $\eta_{ab} \sim 1.5$ [Fig. 2(c)]. This mass anisotropy can be attributed to the underlying anisotropy of the band structures and Fermi surfaces. Figure 2(e) shows that the hole pocket dispersion is much flatter along the X-S (b-axis) direction than the S-R (c-axis) and Y-$X_1$ (a-axis) directions, and the electron pocket dispersion is slightly flatter along the G-Y (b-axis) direction than the Y-T (c-axis) and Y-$X_1$ (a-axis) directions. Moreover, the hole Fermi surfaces are open and possess a spaghetti-like structure that extends along the b direction [Fig. 2(f)]. Therefore, the band mass is naturally expected to be largest along the b-axis, which corresponds to a plasma frequency that is smallest along the b-axis, in agreement with the experimental observations [Fig. 2(c)]. However, we cannot distinguish between the electrons and holes contributions at the present stage. It is well known that Type-II Weyl semimetals are characterized by a significantly tilted Weyl cone with highly anisotropic band dispersions [27, 28]. Both the anisotropic plasma edge and anisotropic Weyl cone are manifestations of the anisotropic electronic structure, which are essentially rooted in the anisotropic atomic structure of $WP_2$.

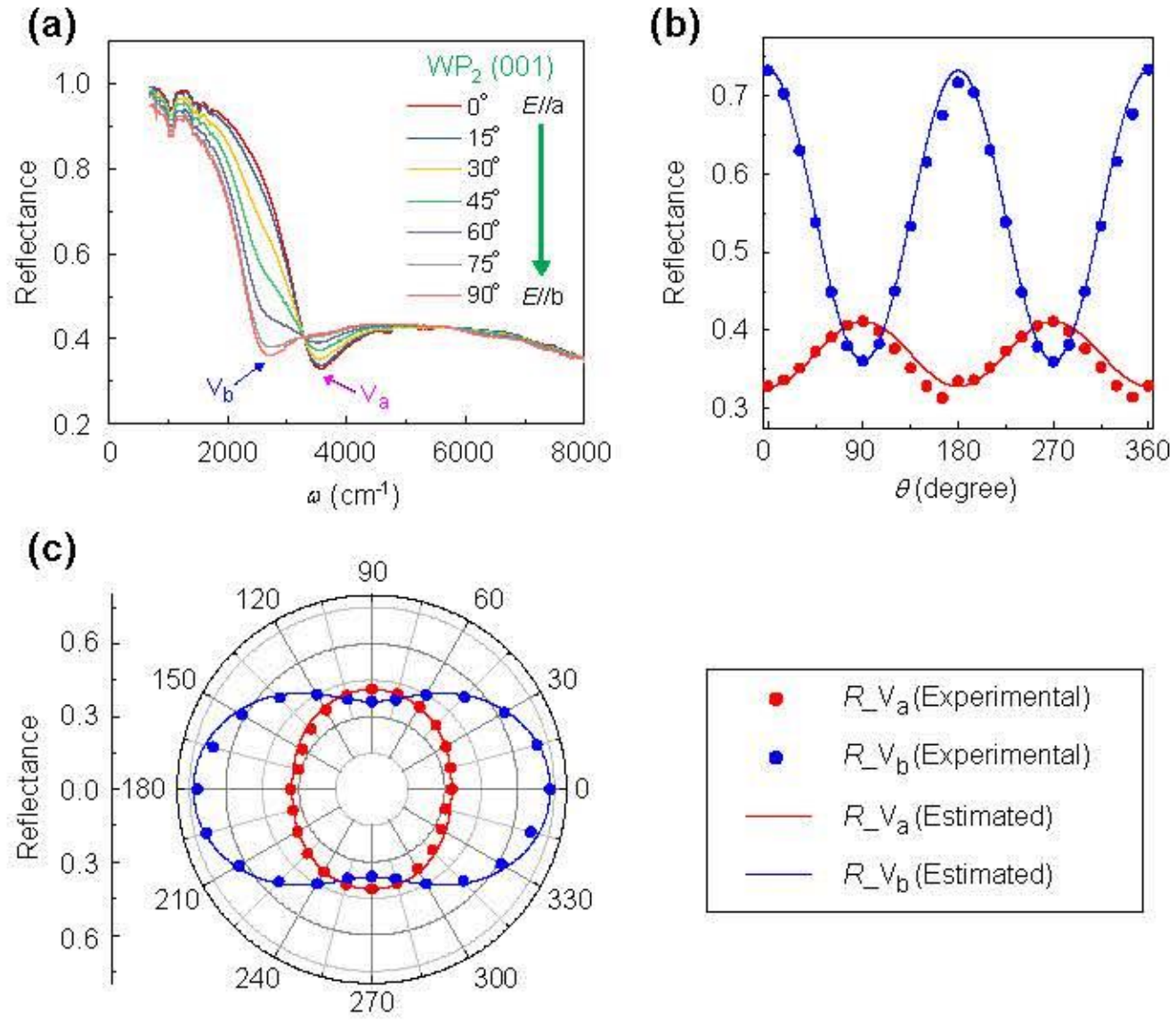

FIG. 3. Polarization-resolved anisotropic reflectance of the $WP_2$ (001) surface. (a) Reflectance spectra at various polarization angle $\theta$s, where $\theta$ denotes the angle between the light polarization direction (direction of electrical field $E$) and the a-axis. Reflectance valleys $V_a$ and $V_b$ are indicated by the arrows. (b) Measured reflectances at the wavenumbers of reflectance valleys $V_a$ ($\sim 3600\ \mathrm{cm}^{-1}$, red circles) and $V_b$ ($\sim 2700\ \mathrm{cm}^{-1}$, blue circles). The corresponding estimated reflectances are depicted as red and blue solid curves, respectively. (c) Polar plot of the measured and estimated reflectances from (b).

According to Fig. 3(a), as the angle ($\theta$) between $E$ and the a-axis increases from 0° to 90°, the plasma reflectance edge of the (001) surface evolves from a single valley ( $V_a$) into double valleys ( $V_a$ and $V_b$), and then finally into a single valley ($V_b$) (see more data for the (001), (010), (021), and (062) surfaces in Figs. S1, S3, and S4, and additional discussion in the SM [25]). At an intermediate polarization angle, the electric field can be decomposed into two orthogonal directions along the a- and b-axes. Therefore, the reflectance can be estimated by the formula $R(\theta) = R(E//a)\cos^2\theta + R(E//b)\sin^2\theta$, where $R(E//$a) and $R(E//$b) are the measured reflectance values at $\theta = 0°$ and $\theta = 90°$, respectively. The estimated reflectances obtained from this formula are nearly identical to experimentally measured values, as shown in Fig. S2 of the

SM [25].

Reflectance anisotropy is more clearly manifested in the $\theta$-dependent reflectances at energies near the reflectance edge. Figures 3(b) and 3(c) show the reflectance as a function of $\theta$ when the wavenumbers are fixed at those of the reflectance valleys $V_a$ ($\sim$ 3600 cm$^{-1}$) and $V_b$ ($\sim$ 2700 cm$^{-1}$). The experimental results reveal a twofold symmetry, which are consistent with the estimated ones obtained from $R(\theta) = R(E//a)\cos^2\theta + R(E//b)\sin^2\theta$. Such polarization-sensitive anisotropic reflectance renders the $WP_2$ an inherent polarization analyzer to detect the light polarization direction. More importantly, by simply tuning the incident light polarization, the reflected light quantity by $WP_2$ was effectively controlled. For example, the reflectance of $WP_2$ at $\sim$ 2700 cm$^{-1}$ declines from $\sim$ 75% to $\sim$ 35% as the polarization angle $\theta$ rotates from 0° (E//a) to 90° (E//b) [see the blue curve in Fig. 3(b)]. Such a large modulation of the reflectance (or reflected light quantity) renders $WP_2$ itself a new concept of prototypical polarization-controlled optical switch based on the mechanism of the polarization-dependent anisotropic plasma edge (see more discussions in the Supplemental note 3 and note 4 of the SM [25]).

Next, we investigate the temperature dependence of the plasma edge, using the $WP_2$ (021) surface with $E//a$ as an example. As shown in Fig. 4(a), the plasma edges are nearly identical across a range of temperatures spanning 300 K down to 5 K. The electron and hole pockets and the corresponding carrier density of $WP_2$ are relatively large, thereby tuning the plasma edge into an energy range of 0.2-0.5 eV. This energy range of plasma edge is much higher than that of previously investigated typical semimetals (e.g., < 0.08 eV for bismuth [15, 16] and $WTe_2$ [19, 20]). Furthermore, the sharp plasma reflectance edges of the $WP_2$ are located away from the interband transitions [Fig. 2(d)]. Both factors may render the plasma edge insensitive to temperature, which facilitates the manipulation and utilization of the reflected light spectrum of $WP_2$ over a wide temperature range. As a result, $WP_2$ is a promising material for application toward multifunctional (e.g., polarization detection and control) photonic and optoelectronic devices. It is noted that other topological semimetals, such as nodal-line semimetals, can also possess intrinsically large anisotropy and large Fermi surfaces at the same time [29, 30], which may also promise anisotropic plasma mirror behavior and corresponding applications and thus extend our findings toward even broader prospects.

### C. Polarization-controlled optical switch application of $WP_2$

Now, we demonstrate the polarization-controlled optical switch function of $WP_2$ by using the photoconductive effect of the HgCdTe semiconductor. As depicted in Fig. 4(b), a linearly polarized laser is rotated by a $\lambda/2$-wave plate and subsequently reflected by the $WP_2$ (001) surface before it finally arrives at a target HgCdTe thin film (see more details in the SM [25]). The adopted laser wavelength is 4.57 $\mu$m, corresponding to a wavenumber of 2188 cm$^{-1}$ (see more discussions in the Supplemental note 5 of the SM [25]). Here we define the relative change of the HgCdTe resistance as $\Delta R/R = (R_1 - R_2)/R_1$, where $R_1$ and $R_2$ are the resistances before and after illumination, respectively. Fig. 4(c) depicts $\Delta R/R$ as a function of $\theta$ with different laser powers. It is evident that $\Delta R/R$ oscillates with $\theta$, reaches a maximum near 0° and 180°, and declines to a minimum near 90° and 270°. Interestingly, as shown in Fig. 4(d), the period and phase of such $\Delta R/R$ oscillations are in excellent agreement with those of the reflectance oscillations of the $WP_2$ (001) surface. Moreover, no periodic oscillations of the $\Delta R/R$ are observed when the $WP_2$ is replaced by a gold film. Therefore, the $\Delta R/R$ oscillations of the HgCdTe can be unambiguously attributed to the polarization-dependent plasma reflectance edge of $WP_2$. The consistence between the resistance oscillations of HgCdTe and reflectance oscillations of $WP_2$ in turn validate the optical switch function of $WP_2$. We emphasize that $WP_2$ itself, instead of HgCdTe, functions as an optical switch, and the present experiment is a prototypical example rather than the optimal case of the practical device applications.

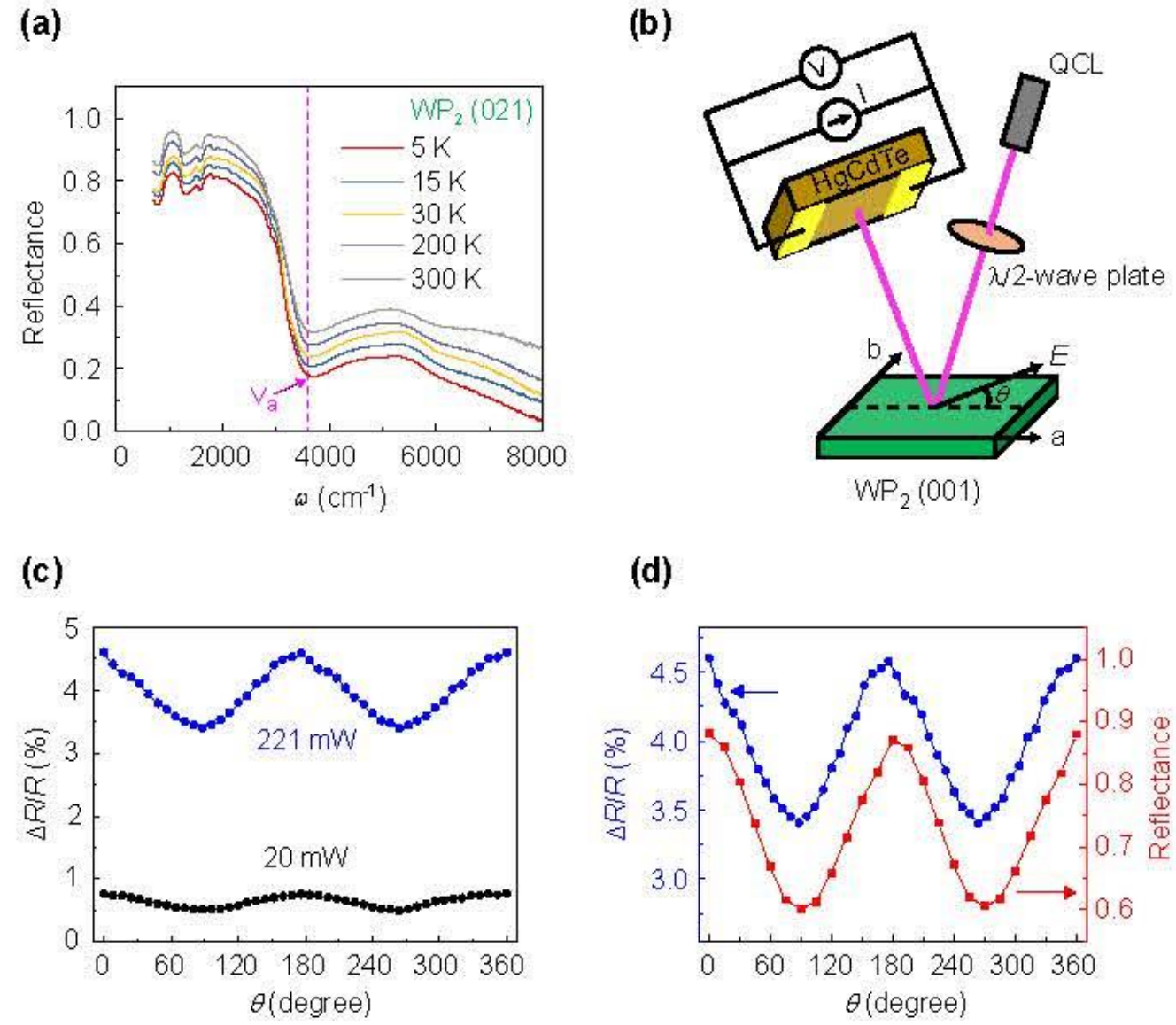

FIG. 4. Polarization-controlled optical switch based on $WP_2$. (a) Reflectance spectra of the (021) surface at various temperatures. The curves are shifted vertically for clarity. The light polarization is along the a-axis. (b) Schematic of the polarization-controlled optical switch apparatus. (c) Relative resistance change $\Delta R/R$ of the HgCdTe thin film as a function of $\theta$, as induced by the reflected light of the $WP_2$ (001) surface. The laser wavelength and wavenumber are 4.57 $\mu$m and 2188 cm$^{-1}$, and the output power of the Quantum Cascade Laser (QCL) is 20 mW (black circles) and 221 mW (blue circles), respectively. (d) The $\Delta R/R$ (blue circles) as a function of $\theta$, where laser power is 221 mW and measurements are taken at room temperature. The measured reflectance (red squares) of the $WP_2$ (001) surface at 2188 cm$^{-1}$ as a function of $\theta$ at room temperature.

The previously investigated optical switches usually consist of complex structures [31]. For example, some opto-mechanical switches are based on micro-electromechanical systems, which require micro/nano manufacture [31]. By contrast, here we have demonstrated a novel and simple polarization-controlled optical switch, which is present in single crystals and can be exploited by taking full advantage of the anisotropic plasma edge of the Weyl semimetal $WP_2$.

## III. CONCLUSION

In summary, we revealed a pronounced anisotropic plasma reflectance edge in the topological Type-II Weyl semimetal $WP_2$, which arises from the corresponding anisotropic electronic structures and is robust against temperature. Moreover, utilizing such polarization-sensitive anisotropic plasma mirror behavior, we achieved a room-temperature $WP_2$-based optical switch, which is effectively controlled by simply tuning the light polarization. The revelation of the anisotropic plasma reflectance edge and the polarization-controlled optical switch not only extend the frontiers of metal optics, but also open the door to new optical applications based on anisotropic topological semimetals.

## ACKNOWLEDGMENTS

This work was supported in part by the National Natural Science Foundation of China (Grant Nos. 11434009, 11774367, U1732148), the National Key R&D Program of China (Grant No. 2017YFA0403600), the Hefei Science Center CAS (Grant No. 2018HSC-UE014), the Anhui Initiative in Quantum Information Technologies (Grant No. AHY170000), the Jiangsu Province Science Foundation for Youth (Grant No. BK20170821), the National Science Foundation of China for Youth (Grant No. 11804160), and the Anhui Provincial Natural Science Foundation (Grant No. 1708085MF136). We thank Hailiang Che, Xuefeng Sun, Han Xu, Zhiyong Lin, Chuansheng Hu, and Qiwei Wang for their experimental support and helpful discussions.

†K.Z., Y.D., and Z.Q. contributed equally to this work.

*Corresponding authors: cgzeng@ustc.edu.cn, weilm203@mail.ustc.edu.cn

**Supplemental Material for:**

# Room-temperature anisotropic plasma mirror and polarization-controlled optical switch based on Type-II Weyl semimetal $WP_2$

Kaixuan Zhang[1,†], Yongping Du[2,†], Zeming Qi[3,†], Bin Cheng[1], Xiaodong Fan[1], Laiming Wei[1,*], Lin Li[1], Dongli Wang[1], Guolin Yu[4], Shuhong Hu[4], Changhong Sun[4], Zhiming Huang[4], Junhao Chu[4], Xiangang Wan[5], and Changgan Zeng[1,*]

[1]*International Center for Quantum Design of Functional Materials, Hefei National Laboratory for Physical Sciences at the Microscale, CAS Key Laboratory of Strongly Coupled Quantum Matter Physics, Department of Physics, and Synergetic Innovation Center of Quantum Information & Quantum Physics, University of Science and Technology of China, Hefei, 230026, China*

[2]*Department of Applied Physics and Institution of Energy and Microstructure, Nanjing University of Science and Technology, Nanjing, Jiangsu 210094, China*

[3]*National Synchrotron Radiation Laboratory, University of Science and Technology of China, Hefei, Anhui 230029, China*

[4]*National Laboratory for Infrared Physics, Shanghai Institute of Technical Physics, Chinese Academy of Sciences, Shanghai 200083, China*

[5]*National Laboratory of Solid State Microstructures, Department of Physics, and Collaborative Innovation Center of Advanced Microstructures, Nanjing University, Nanjing, Jiangsu 210093, China*

[†]K.Z., Y.D., and Z.Q. contributed equally to this work.

[*]Corresponding authors: cgzeng@ustc.edu.cn, weilm203@mail.ustc.edu.cn.

## Methods

**Growth and structure characterization of $WP_2$ single crystals.** High-quality $WP_2$ single crystals were grown by the chemical vapor transport method [1]. P, $WO_3$, and $I_2$ were mixed and sealed in a quartz tube under vacuum, and then the $WP_2$ crystals were grown in a two-zone furnace with a temperature gradient of 1000 ˚C (source) to 900 ˚C (sink) for 10 days. Crystal and surface structures were characterized by X-ray diffraction (XRD), transmission electron microscopy, and scanning tunneling microscopy. As shown in Fig. 1(b), the as-grown $WP_2$ single crystals possess several natural crystal faces parallel to the a-axis including the (010), (062), and (021) surfaces. The optical responses for *E*//a and *E*//c can be measured on the (010) surface, whereas the response for *E*//b cannot be directly measured on these three surfaces. In order to measure the complete optical response of $WP_2$, it is imperative to fabricate other simple crystal faces with low indices, such as the (001) surface, which can be used to measure the optical response for *E*//b. Using X-ray Laue photography, the crystals were carefully oriented and cut into small pieces along the appropriate crystalline axes to obtain the (001) surface (see more details in Supplemental note 1).

**Polarized Fourier-transform infrared spectroscopy measurements.** The mid-infrared reflectance measurements were performed with a Fourier-transform infrared spectrometer (Bruker IFS 66v) on the infrared beamline (BL01B) at the National Synchrotron Radiation Laboratory of China.

**Electronic structures and optical property calculations.** The electronic structures were calculated using the local spin density approximation (LSDA) and the full-potential linearized augmented plane-wave (FP-LAPW) method [2] as implemented in the WIEN2k [3]. The system was assumed to be non-magnetic. The plane-wave cutoff parameter $R_{MT}*K_{max}$ was set to be 7 and a $24\times 24\times 15$ mesh was used for Brillouin-zone sampling during the iteration for self-consistency, both of which guarantee the convergence of total energy (Figs. S5a and S5b). The spin-orbit coupling was treated using the second-order variational procedure. For the Fermi surface calculation of few-layer $WP_2$, due to the large amount of calculation, we were only able to perform the Fermi surfaces calculation up to a thickness of 11 layers within a quarter of Brillouin zone (Fig. S6).

The optical properties and the density of states calculations were performed using OPTIC and DOS modules of WIEN2k. The optical calculations required a much finer mesh of k points, so a grid of $48\times 48\times 30$ was adopted for optical properties calculations. In fact, ~10000 k points is sufficient for the optical calculations (Fig. S7). The scattering rate $\Gamma$ was adjusted to be 0.05 eV (~403 $cm^{-1}$) to match the experimental value of ~300-478 $cm^{-1}$ (see Table S1). We would like to mention that the reflectance change induced by a small variation of $\Gamma$ is negligible. For example, the calculated reflectance spectra with $\Gamma$=0.05 eV (~403 $cm^{-1}$) and $\Gamma$=0.074 eV (~600 $cm^{-1}$) are almost identical, as shown in Fig. S8.

**Fabrication and measurement of the polarization-controlled optical switch application of $WP_2$.** The linearly polarized laser was produced by a 4.57-μm Quantum Cascade Laser (QCL). The light polarization direction was continuously tuned by rotating the λ/2-wave plate. $Hg_{1-x}Cd_xTe$ (x = 0.22) film was grown on a CdZnTe substrate by liquid phase epitaxy [4] and subsequently polished to obtain a flat surface. Indium electrodes were burned onto the HgCdTe to form contacts. The bandgap of $Hg_{1-x}Cd_xTe$ is ~ 0.2 eV for x = 0.22 at 300 K [5]. The 4.57-μm laser illumination could excite electrons from the valence band of HgCdTe to the conduction band, thereby increasing the conductivity in response to the number of incident photons. The photoconductivity of HgCdTe was measured by a two-terminal method with a constant current (1 mA) at room temperature.

## Supplemental notes

### 1. Fabrication of the $WP_2$ (001) surface

Using X-ray Laue photography, the $WP_2$ crystals were carefully oriented and cut into small pieces along the appropriate crystalline axes to obtain the (001) surface. First, we cut the as-grown crystal perpendicular to the a-axis to obtain the bc plane. Then we determined the direction of b- and c-axes on the bc plane via Laue photography. Finally, the (001) surface was obtained by cutting the crystal parallel to the a- and b-axes (i.e., perpendicular to the c-axis). The resulting surface was subsequently polished with finely ground $Al_2O_3$ powder.

### 2. Polarization-resolved anisotropic reflectance of the (021) and (062) surfaces

Polarization-resolved anisotropic reflectance was also measured on high Miller index surfaces, e.g., the (021) and (062) surfaces. Figures S3(a) and S4(a) depict the schematic of the measurement configuration. The black rectangles denote the (021) and (062) surfaces, and $\theta$ denotes the angle between the light polarization direction (direction of electrical field $E$) and the a-axis. Figures S3(b) and S4(b) show the corresponding XRD patterns of the high-quality single crystals, each of which has a small full width at half maximum (FWHM). Figures S3(c) and S4(c) show that when the light polarization is along the a-axis ($\theta = 0^\circ$), the reflectance exhibits a plasma reflectance edge with a single valley $V_a$ at ~3600 $cm^{-1}$. This value is almost identical to the measured $V_a$ of the (001) and (010) surfaces [Figs. 2(a) and 2(b)]. Moreover, when the light polarization lies in the bc plane ($\theta = 90^\circ$), the reflectance develops a plasma reflectance edge with two valleys $V_b$ at ~2700 cm$^{-1}$ and $V_c$ at ~4000 cm$^{-1}$, which are nearly identical to the measured $V_b$ and $V_c$ of the (001) and (010) surfaces [Figs. 2(a) and 2(b)]. Although $V_a$, $V_b$, and $V_c$ are expected to coexist at certain $\theta$s between 0˚ and 90˚, the valleys $V_a$ and $V_c$ are too close to be well-resolved at these $\theta$s.

3. **Possible optimizations for the $WP_2$ optical switch**

For the ideal plasma reflectance edge, the reflectance will sharply drop exactly from 1 to 0 at the plasma edge, resulting in an ideal optical switch without any leakage. Here in practice, the present optical switch by one single $WP_2$ crystal indeed has a leakage reflectance. However, such a leakage can be effectively suppressed by adopting $WP_2$ crystals in series or adding specific optical cutoffs. Practical applications require much more efforts, including the above optimizing processes or exploring new materials with better performances in the future.

4. **Potential miniaturization for the $WP_2$ optical switch**

For practical uses, the size, in particular, the thickness of the material is preferred to be miniaturized. Therefore, it is useful to highlight the application prospects by obtaining the thickness dependence of the reflectance anisotropy or by estimating the minimum thickness that this anisotropy in reflectance contrast remains similar to bulk values.

Unfortunately, it is practically impossible to exfoliate such non-van-der-Waals crystals into nanoflakes with different nanoscale thicknesses, and we are unable to grow $WP_2$ single-crystalline thin films at the present stage. Therefore, we cannot experimentally obtain the thickness dependence of the reflectance anisotropy. Nevertheless, we could perform related theoretical calculations (see more details in the Methods section). Figure S6 shows the calculated electronic structures (Fermi surfaces) on the (001) surface with different thicknesses of $WP_2$. It can be seen that when the thickness is above 7 layers (3.5 nm), the Fermi surfaces start to exhibit a pattern quite similar to that of bulk $WP_2$. Since the anisotropic reflectance roots in the anisotropic electronic structures, the anisotropy in reflectance contrast is naturally expected to remain similar to bulk values for $WP_2$ with a thickness above 3.5 nm.

On the other hand, the skin depth is roughly 100-400 nm (Fig. S9d) within the frequency range of 1600-4000 $cm^{-1}$. For practical applications, the thickness of $WP_2$ is better to be larger than the order of the skin depth, i.e., 100 nm, which is roughly 29 times that of 3.5 nm (3.5 nm corresponds to the thickness of 7 layers shown in Fig. S6). For such a thick $WP_2$, the anisotropy in reflectance contrast should remain similar to bulk values.

## 5. **The adopted laser wavelength**

It is noted that the contrast of anisotropic reflectance reaches its maximum at the plasma edge (~2600 $cm^{-1}$), i.e., the reflectance declines from 77% ($E$//a) to 37% ($E$//b) with a decrease ~40%. At the present stage, however, we do not have a light source with a wavenumber of ~2600 $cm^{-1}$, and the most substitutable light source we have is a mid-infrared laser with a wavenumber of ~2188 $cm^{-1}$ (the corresponding wavelength is 4.57 μm). Nevertheless, the contrast of anisotropic reflectance is still considerably large at ~2188 $cm^{-1}$, i.e., the reflectance declines from 88% (*E*//a) to 60% (*E*//b) with a decrease of ~28%. More importantly, the experiment performed at ~2188 $cm^{-1}$ indicated that the photo-induced resistance oscillations of HgCdTe are well consistent with the reflectance oscillations of $WP_2$ (001) surface, as shown in Fig. 4(d) in the main text. Therefore, the present experiment carried out at ~2188 $cm^{-1}$ is sufficient to demonstrate the optical switch function of $WP_2$.

**Supplemental figures**

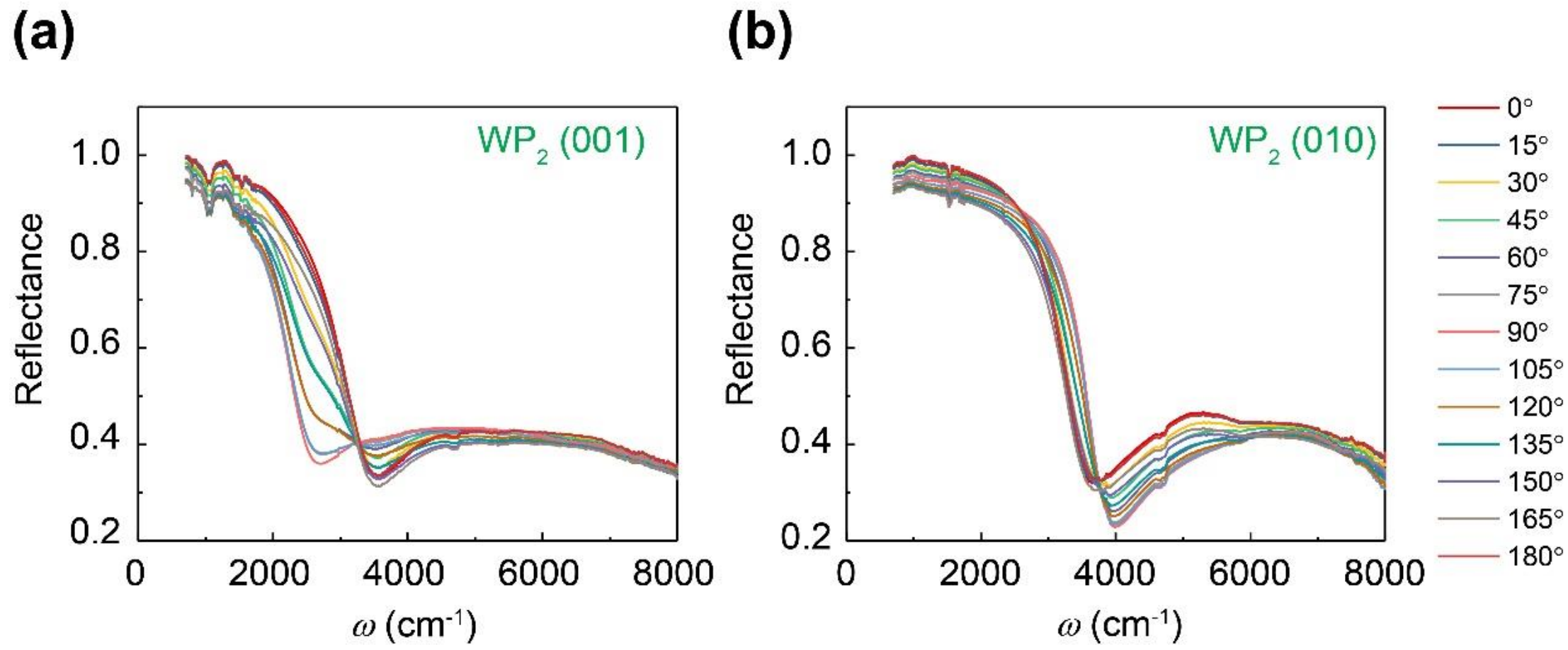

Figure S1. Measured reflectance from $\theta = 0^\circ$ to $\theta = 180^\circ$ for the (001) (a) and (010) (b) surfaces, respectively. $\theta$ denotes the angle between light polarization direction and the a-axis.

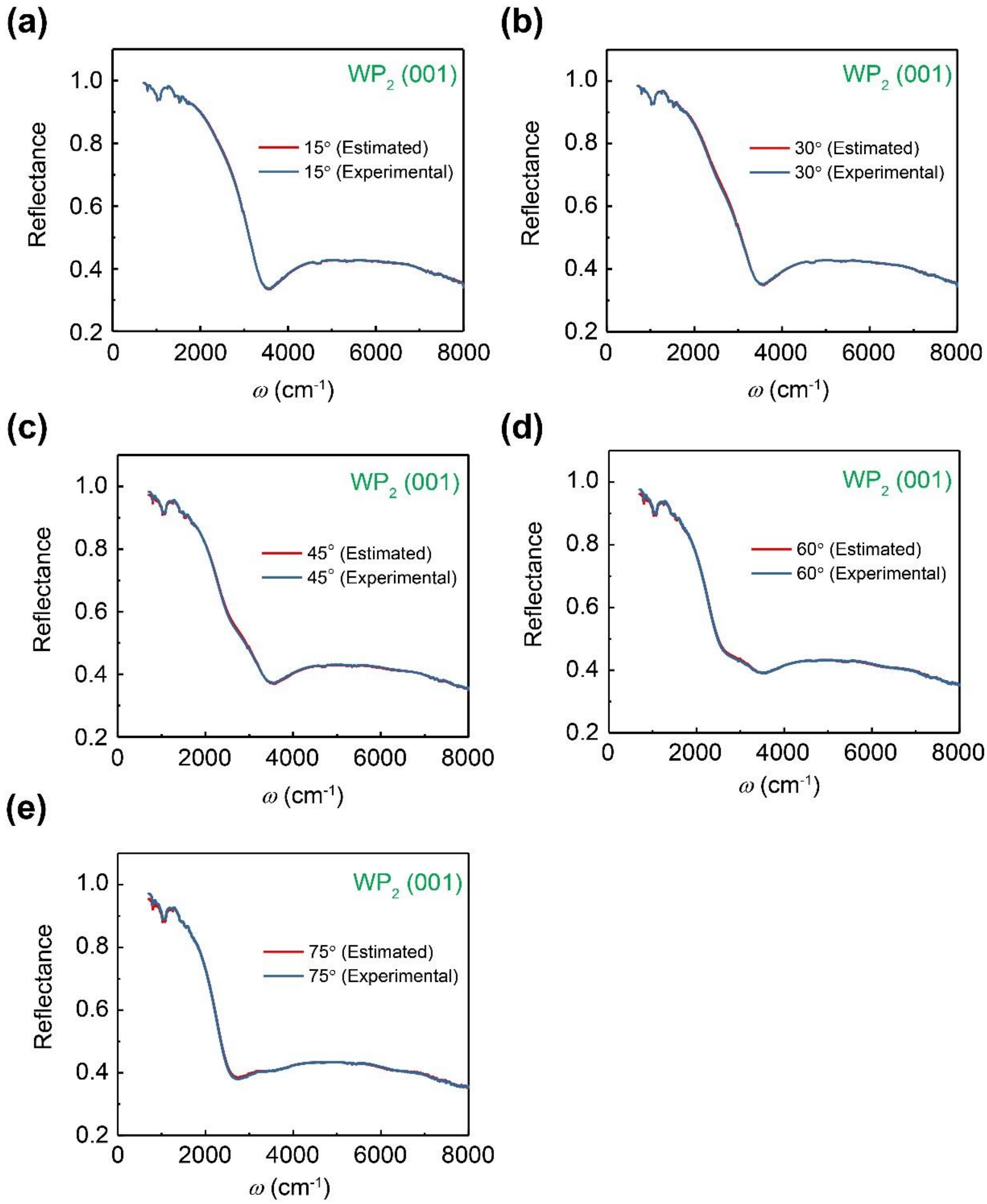

Figure S2. Experimental (blue) and estimated (red) reflectance of the $WP_2$ (001) surface at $\theta = 15^\circ$ (a), $30^\circ$ (b), $45^\circ$ (c), $60^\circ$ (d), and $75^\circ$ (e), respectively. The estimations are based on the formula $R(\theta) = R(E//a)\cos^2\theta + R(E//b)\sin^2\theta$, where $R(E//a)$ and $R(E//b)$ are the measured reflectances at $\theta = 0^\circ$ and $\theta = 90^\circ$, respectively.

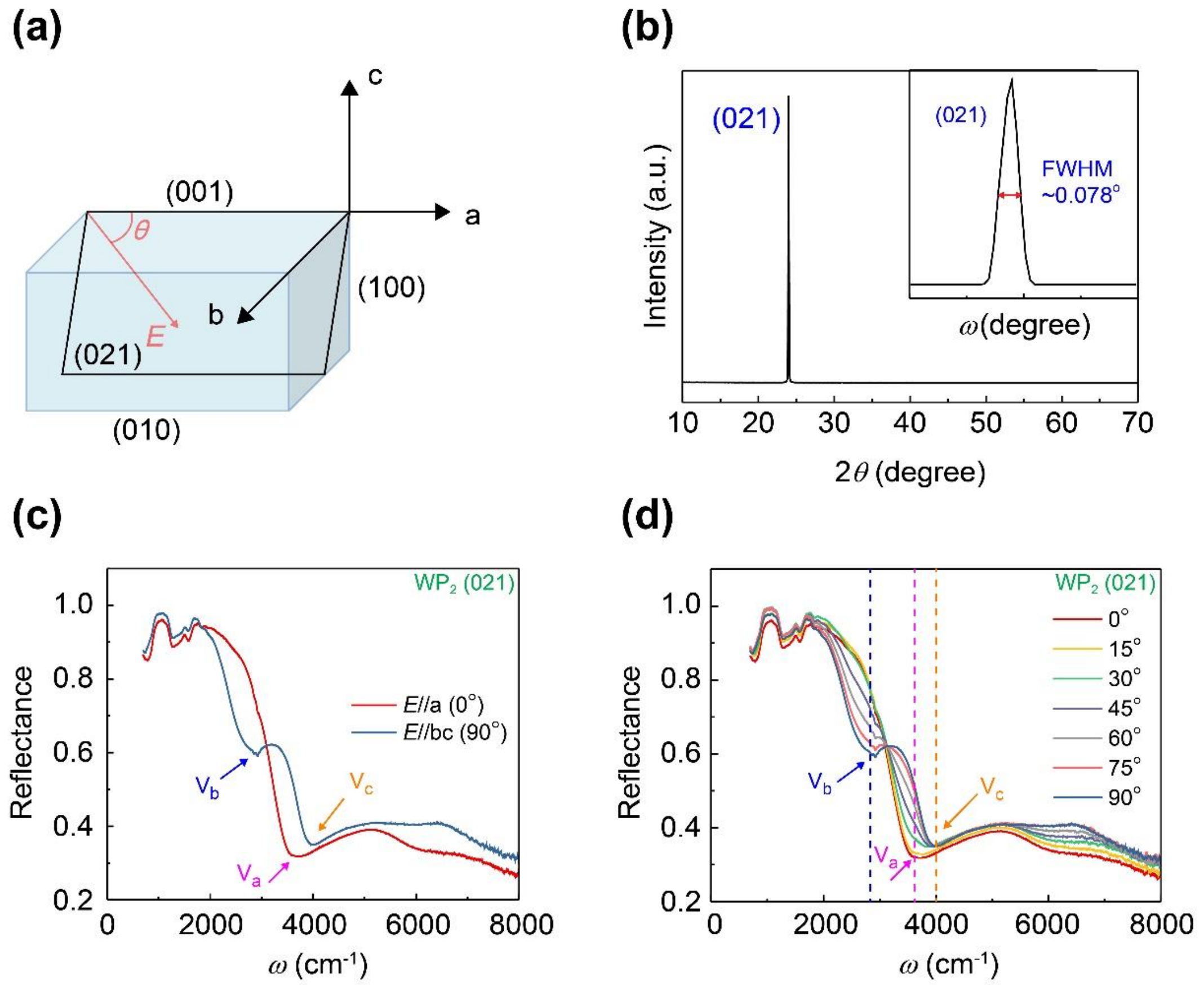

Figure S3. Polarization-resolved anisotropic reflectance of the $WP_2$ (021) surface. (a) Schematic of the measurement configuration. (b) XRD pattern. The inset shows the rocking curve. (c) Reflectance measured at $\theta = 0^\circ$ (*E*//a, red solid curve) and $\theta = 90^\circ$ (*E*//bc plane, blue solid curve). (d) Reflectance at various polarization angle $\theta$s. The pink, blue, and yellow arrows and lines indicate the reflectance valleys $V_a$, $V_b$, and $V_c$, respectively.

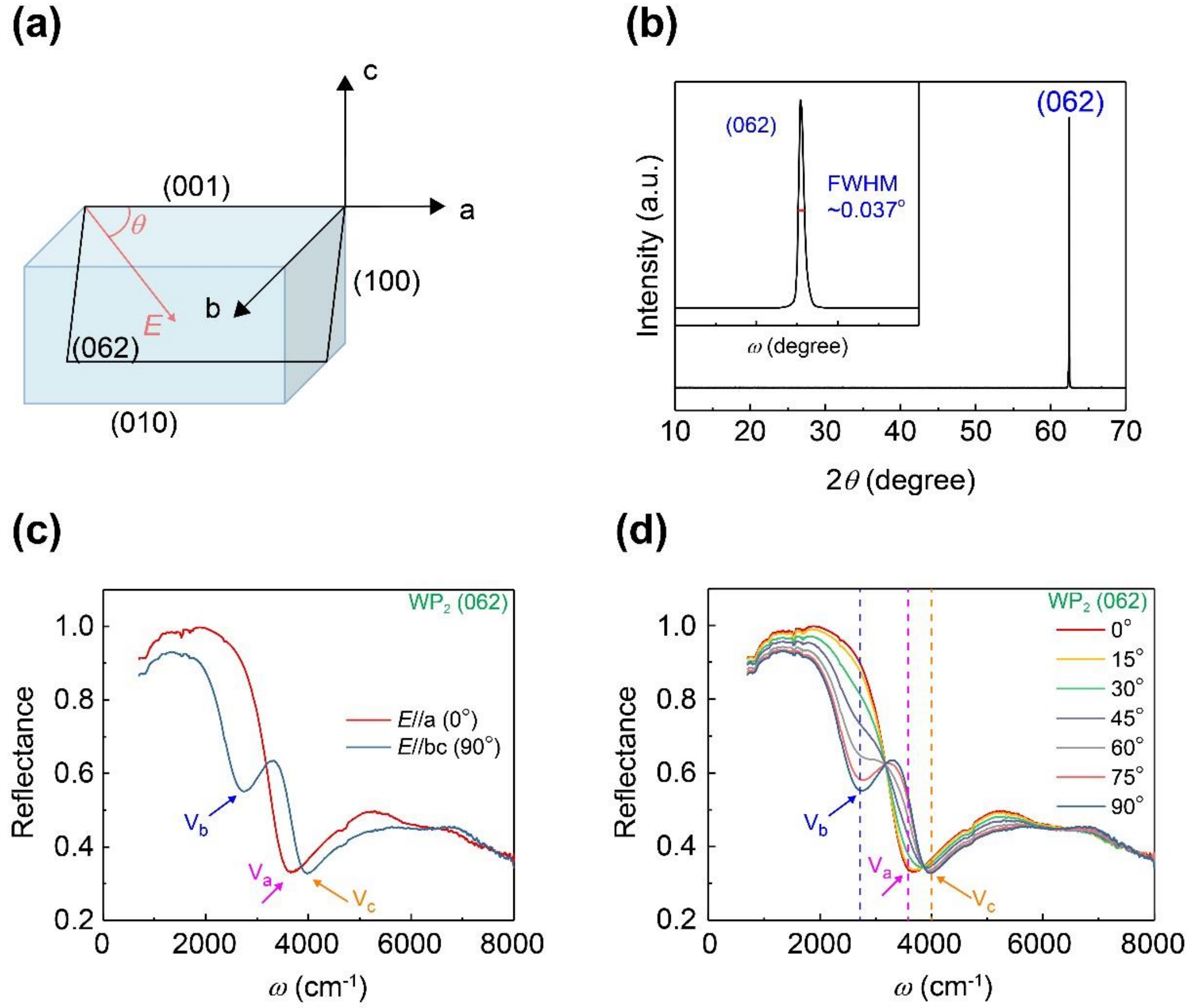

Figure S4. Polarization-resolved anisotropic reflectance of the $WP_2$ (062) surface. (a) Schematic of the measurement configuration. (b) XRD pattern. The inset shows the rocking curve. (c) Reflectance measured at $\theta = 0^\circ$ ($E$//a, red solid curve) and $\theta = 90^\circ$ ($E$//bc plane, blue solid curve). (d) Reflectance at various polarization angle $\theta$s. The pink, blue, and yellow arrows and lines indicate the reflectance valleys $V_a$, $V_b$, and $V_c$, respectively.

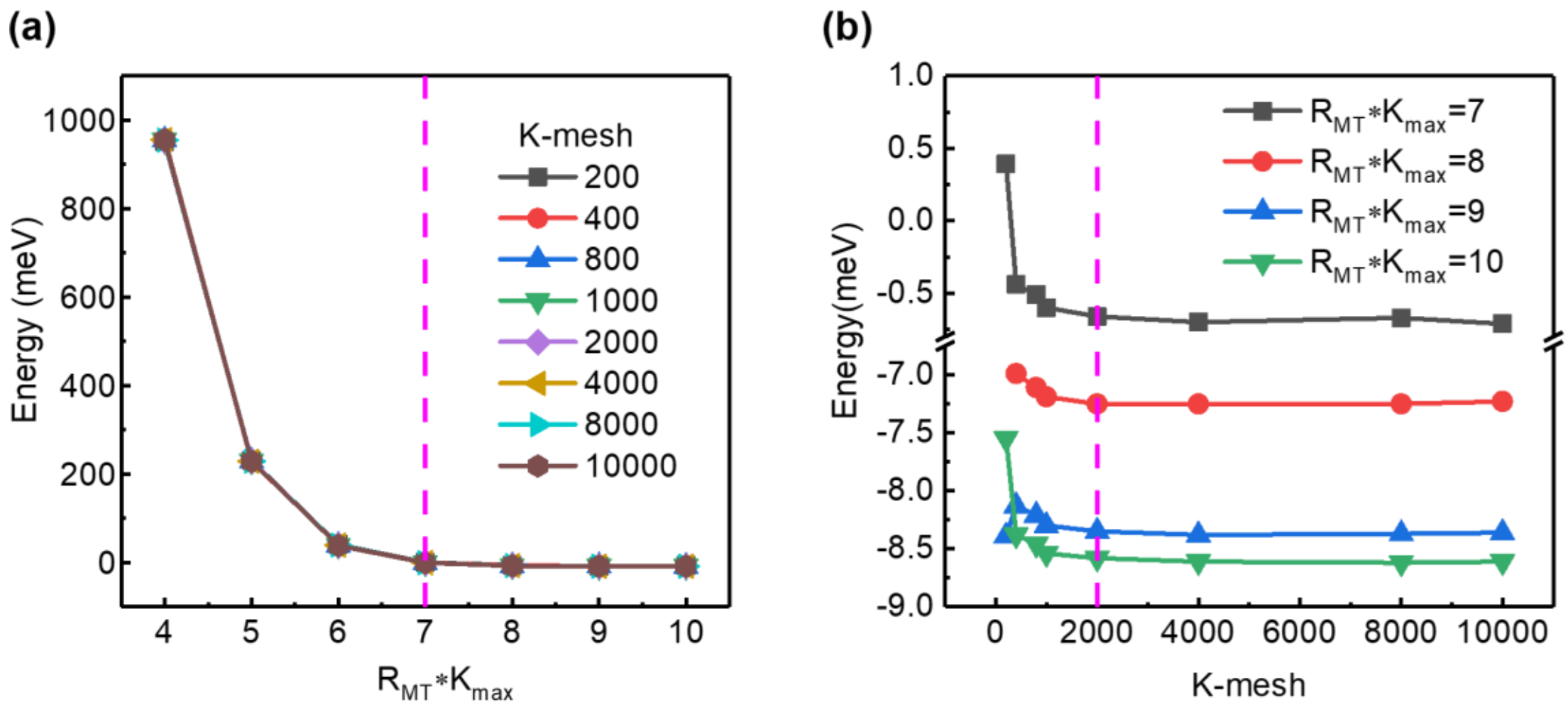

Figure S5. (a) Total energy as a function of $R_{MT}$*$K_{max}$ at various K-mesh values. (b) Total energy as a function of K-mesh at different $R_{MT}$*$K_{max}$ values. The pink dashed lines indicate the position where the total energy starts to converge.

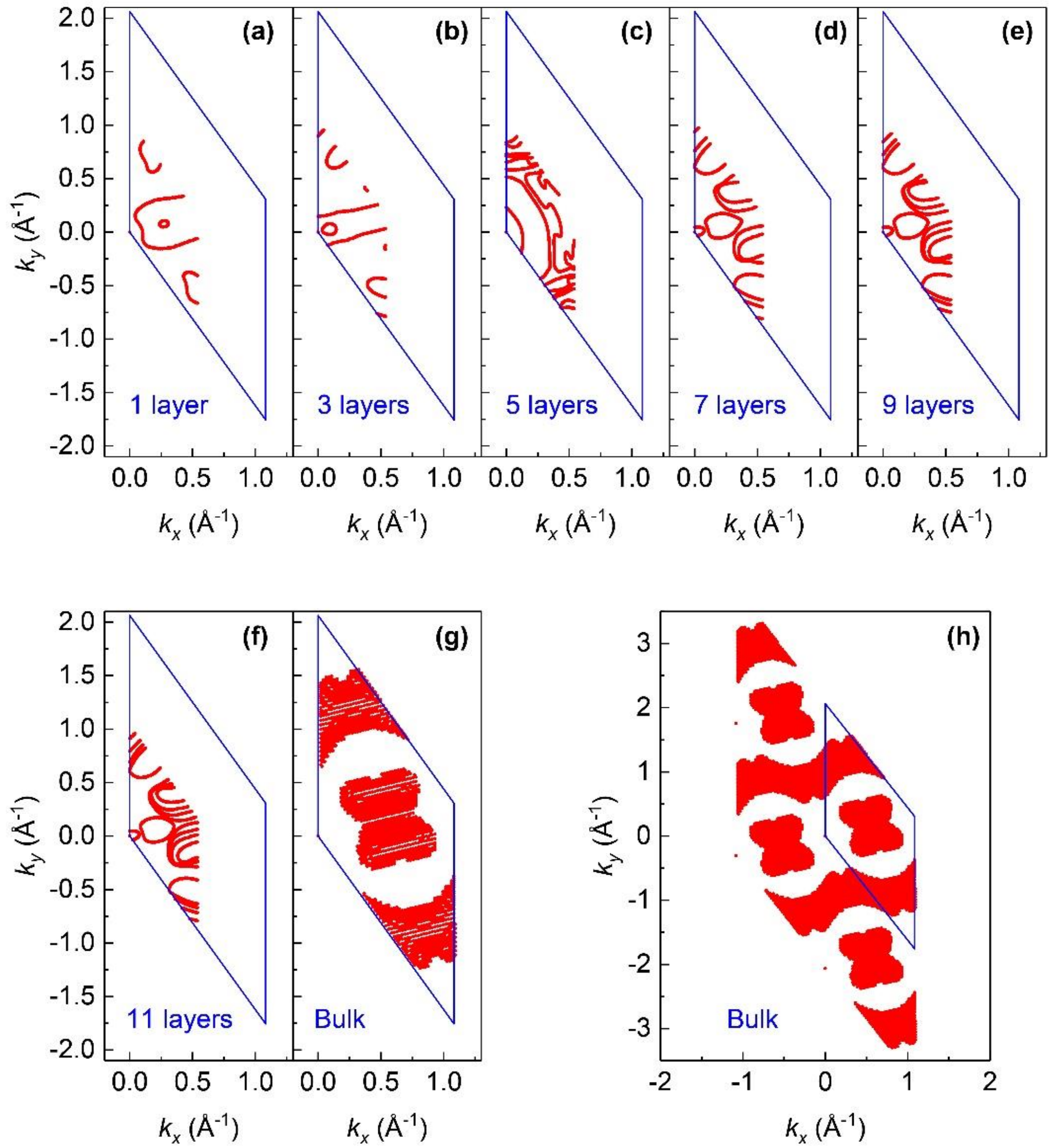

Figure S6. Calculated Fermi surfaces of few-layer and bulk $WP_2$. (a-f) Calculated Fermi surfaces on the $k_x$-$k_y$ plane for $WP_2$ with various thicknesses: 1 (a), 3 (b), 5 (c), 7 (d), 9 (e),11(f) layers, respectively. The blue lines indicate the Brillouin zone boundaries. (g-h) Projection of the Fermi surfaces for bulk $WP_2$ on the $k_x$-$k_y$ plane. The blue lines indicate the Brillouin zone boundaries.

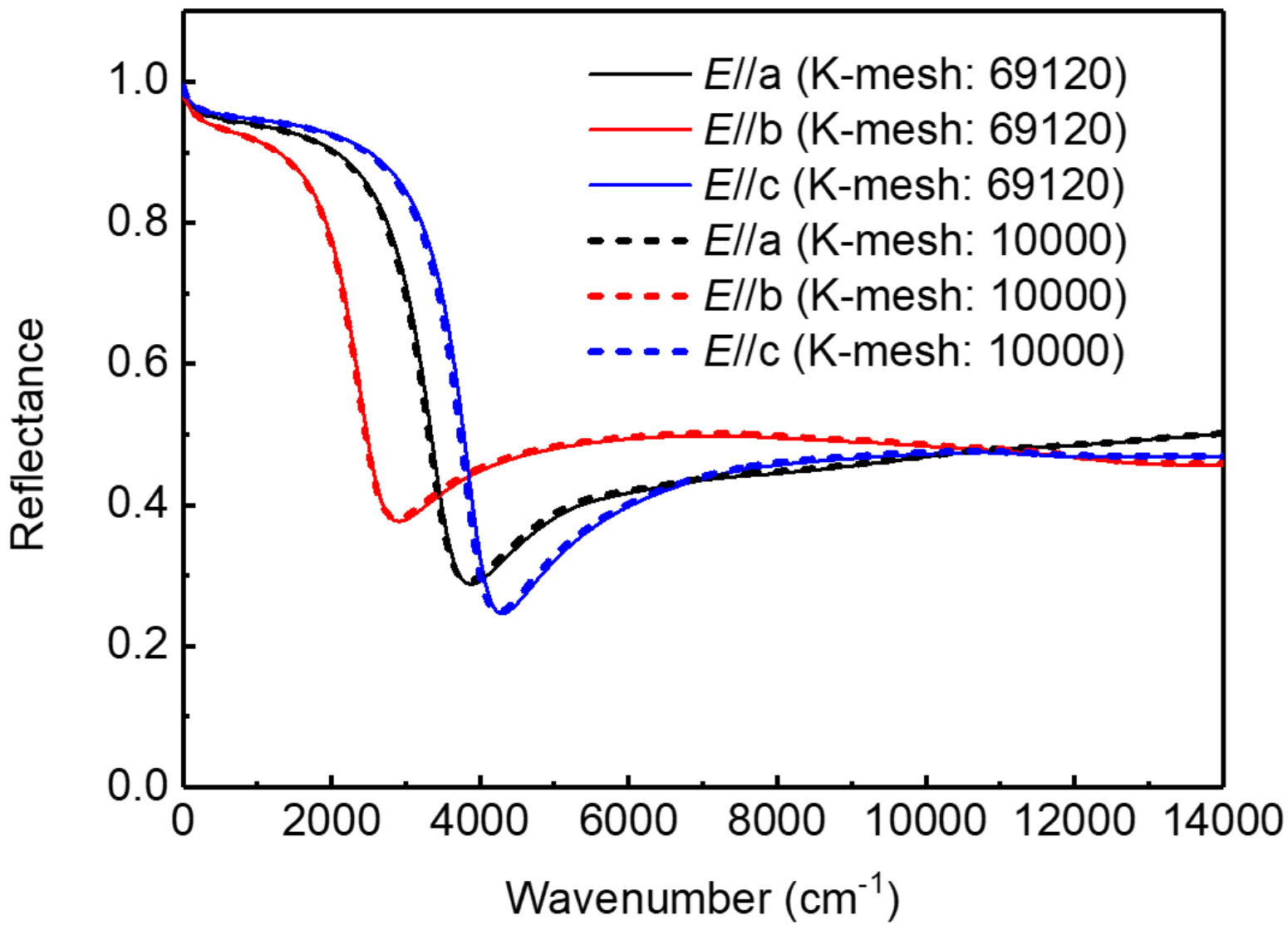

Figure S7. Calculated reflectance with 10000 k points (dashed curves) and 69120 k points (solid curves), respectively.

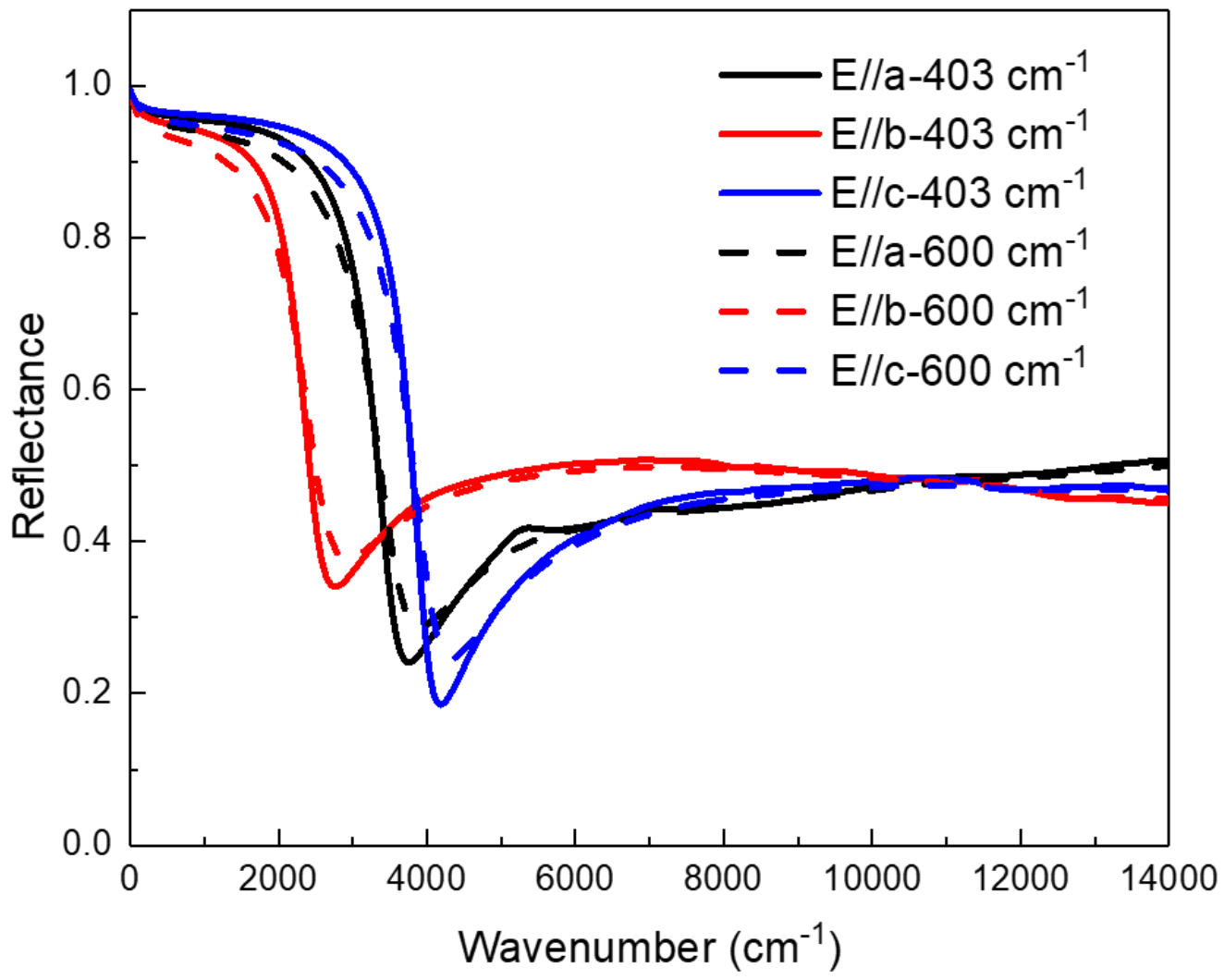

Figure S8. Calculated reflectance for *E*//a, *E*//b, and *E*//c, with the scattering rate adopted to be 403 $cm^{-1}$ (solid) and 600 $cm^{-1}$ (dashed), respectively.

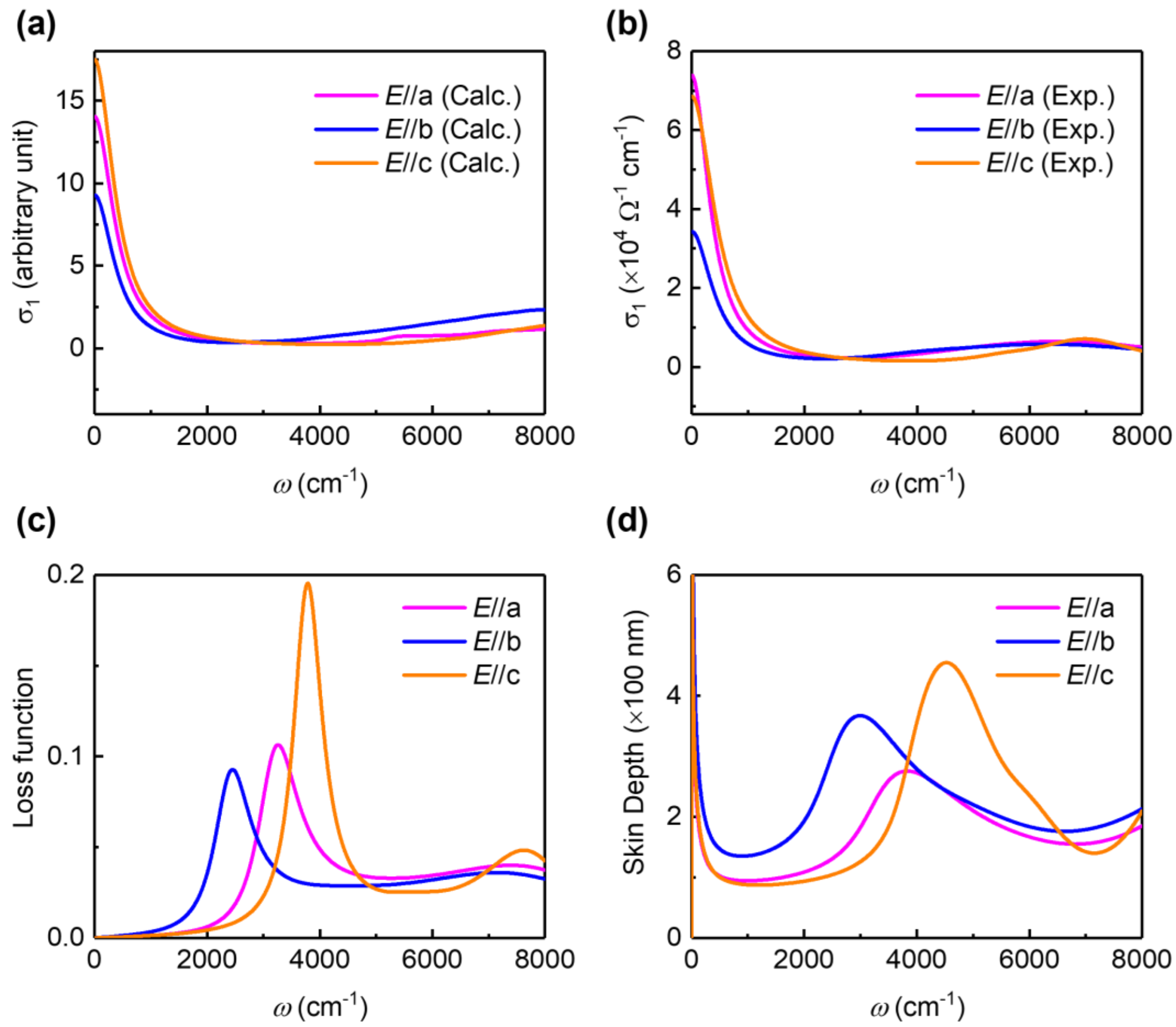

Figure S9. Optical properties of $WP_2$ single crystals. (a-b) Calculated (a) and experimental (b) optical conductivity $\sigma_1$ for *E*//a, *E*//b, and *E*//c, respectively. It is noted that the experimental anisotropic optical conductivity spectra are consistent with the calculated ones. (c) Loss function extracted by fitting the experimental data, which is relatively small. (d) Skin depth extracted by fitting the experimental data.

## Supplemental table

| | $\varepsilon_\infty$ | $\omega_{p,j}$ (cm$^{-1}$) | $1/\tau_j$ (cm$^{-1}$) | $\omega_{0,k}$ (cm$^{-1}$) | $\Omega_{p,k}$ (cm$^{-1}$) | $\gamma_k$ (cm$^{-1}$) |
|---|---|---|---|---|---|---|
| E//a | 22.1 | 16492 | 404 | 6915 | 16070 | 4554 |
| | | 8750 | 300 | 5319 | 10929 | 4004 |
| E//b | 20.0 | 12288 | 481 | 4192 | 7631 | 2880 |
| | | 6404 | 350 | 6536 | 18118 | 5224 |
| E//c | 22.5 | 18233 | 478 | 5602 | 5018 | 1553 |
| | | 8900 | 478 | 7015 | 13110 | 2132 |

Table S1. The fitting parameters for *E*//a, *E*//b, *E*//c, respectively. $\varepsilon_\infty$ is the permittivity at high frequency, $\omega_{p,j}$ are the free carrier plasma frequencies for electrons and holes, $\tau_j$ are the free carrier scattering times for electrons and holes, $\Omega_{p,k}$ are the oscillator strengths for phonons and interband electronic transitions, $\omega_{0,k}$ are the phonon and interband transition frequencies, and $\gamma_k$ is the width of the corresponding transition. The fitting formula is listed below:

$$\varepsilon(\omega) = \varepsilon_\infty - \sum_{j=1}^{2} \frac{\omega_{p,j}^2}{\omega^2 + i\omega/\tau_j} + \sum_k \frac{\Omega_{p,k}^2}{\omega_{0,k}^2 - \omega^2 - i\omega\gamma_k}$$